# Multi-contact Phase Change Toggle Logic Device Utilizing Thermal Crosstalk

Raihan Sayeed Khan[1], Nadim H. Kan'an[1], Jake Scoggin[1], Helena Silva[1] and Ali Gokirmak[1]

[1]*University of Connecticut, Storrs, CT 06269, USA*

*Email: raihan.khan@uconn.edu / Phone: +1 860 208 9253*

Phase change memory (PCM) is an emerging high speed, high density, high endurance, and scalable non-volatile memory technology which utilizes the large resistivity contrast between the amorphous and crystalline phases of chalcogenide materials such as $Ge_2Sb_2Te_5$ (GST). In addition to being used as a standalone memory, there has been a growing interest in integration of PCM devices on top of the CMOS layer for computation in memory and neuromorphic computing. The large CMOS overhead for memory controllers is a limiting factor for this purpose. Transferring functionality like routing, multiplexing, and logic to the memory layer can substantially reduce the CMOS overhead, making it possible to integrate 100s of GB of PCM storage on top of a conventional CPU [1], [2].

In this work, we present computational analysis of a phase change device concept that can perform toggle operations. The toggle functionality is achieved using two physical mechanisms: (i) isolation of different read contacts due to amorphization between different write contact pairs, and (ii) thermal cross-talk between a molten region and a previously amorphized region. Phase-change devices with six contacts can be implemented as toggle flip-flops, multiplexer, or demultiplexer when interfaced with CMOS transistors. Here, we demonstrate the operation of the device as a toggle flip-flop with 5 transistors (Fig. 1a), requiring ~50% of the footprint compared to conventional CMOS alternatives, with the added advantage of non-volatility. The fabrication steps (Fig. 1b-h) for the proposed device are compatible with CMOS back-end-of-line integration.

We have demonstrated the device concept using our finite element framework in COMSOL Multiphysics that can simultaneously capture amorphization-crystallization dynamics including nucleation and growth, and electro-thermal phenomena [3]–[5]. Phase change is modeled by solving a rate equation to track the evolution of the vector field crystal density ($\overrightarrow{CD}$) with $|CD| = CD_1+CD_2 = 1$ or 0 representing the crystalline or amorphous phase, respectively, and individual components ($CD_1$ and $CD_2$) representing the 2D grain orientation. Current continuity and heat transfer equations are solved self-consistently.

A write pulse ($V_{write}$ high) is required to initialize the device to one of the two configurations. During the initialization pulse, one of the paths ($W_{1\text{-}3}$ in Fig. 2b) draws a progressively larger proportion of the current and eventually melts due to thermal runaway. After the pulse is terminated, the path $W_{1\text{-}3}$ is amorphized, blocking the read path $R_{1\text{-}3}$ (Fig. 2c). This results in a low voltage ($V_{low}$ ~ 1 mV for $R_L$=10 kΩ) at Q and a high voltage ($V_{high}$ ~ 50 mV for $R_L$=10 kΩ) at Q′ during a read operation. Applying a subsequent write pulse, the alternative path ($W_{1\text{-}2}$) draws most of the current and eventually melts because $W_{1\text{-}3}$ is initially amorphous (Fig. 2d). As $W_{1\text{-}2}$ melts, $W_{1\text{-}3}$ heats above the crystallization temperature and the device cools to a toggled state (Fig. 2e). Fig. 3 shows the write currents and read voltages during the initialization pulse and two consecutive write pulses showing toggle operation. The voltages applied to the gates of the write and read FETs are 3 V and 0.5 V; power consumed during read and write operations are 585 µW and 0.25 µW for GST thickness of 20 nm. The node Q and Q′ can be connected to a comparator or amplifier to achieve rail-to-rail voltage. Following the write pulse, the amorphous resistivity is low due to increased temperature, thus the ratio of output voltages is smaller for read pulses applied shortly after the write pulse and it increases with time (Fig. 4). Also, as the value of $R_L$ increases, the output voltages will increase, but the ratio of high to low output voltage will decrease (Fig. 4 inset). Therefore, the sensitivity of the comparator will guide the choice of resistors connected to the read nFETs.

The speed of the device itself is determined by the distance between write contacts (shorter distance will result in faster re-crystallization, thus higher speed), placement of thermal anchor (the closer the thermal anchors, the less time it will take to cool down to room temperature at the cost of additional power), and the size of the write FETs (larger FET will provide more current at the cost of increased footprint). The proposed six contact device is slower than a conventional CMOS implementation (~.3 ns for CMOS vs. ~10 ns for proposed device) but reduces the footprint by ~50% and can reduce energy consumption significantly in applications with infrequent writes due to non-volatility.

This work is supported by NSF under award # ECCS 1711626.

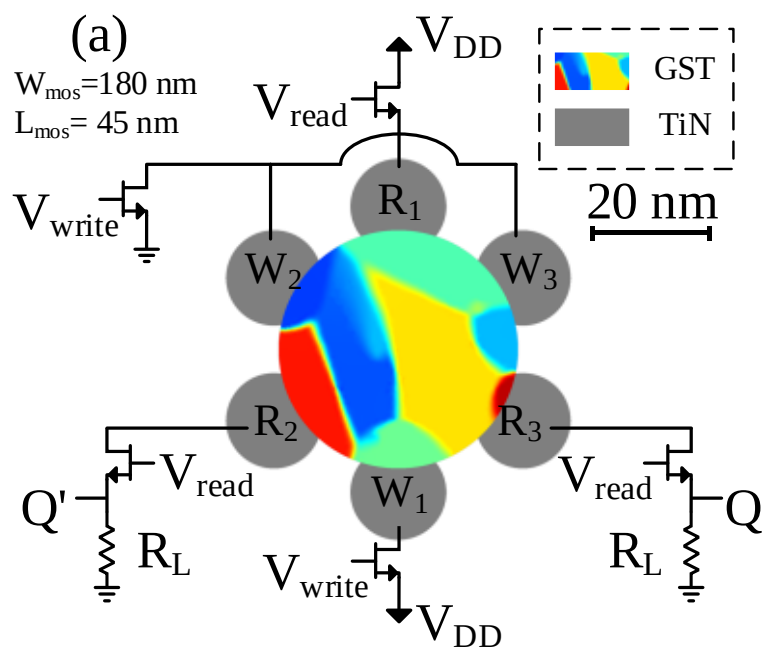

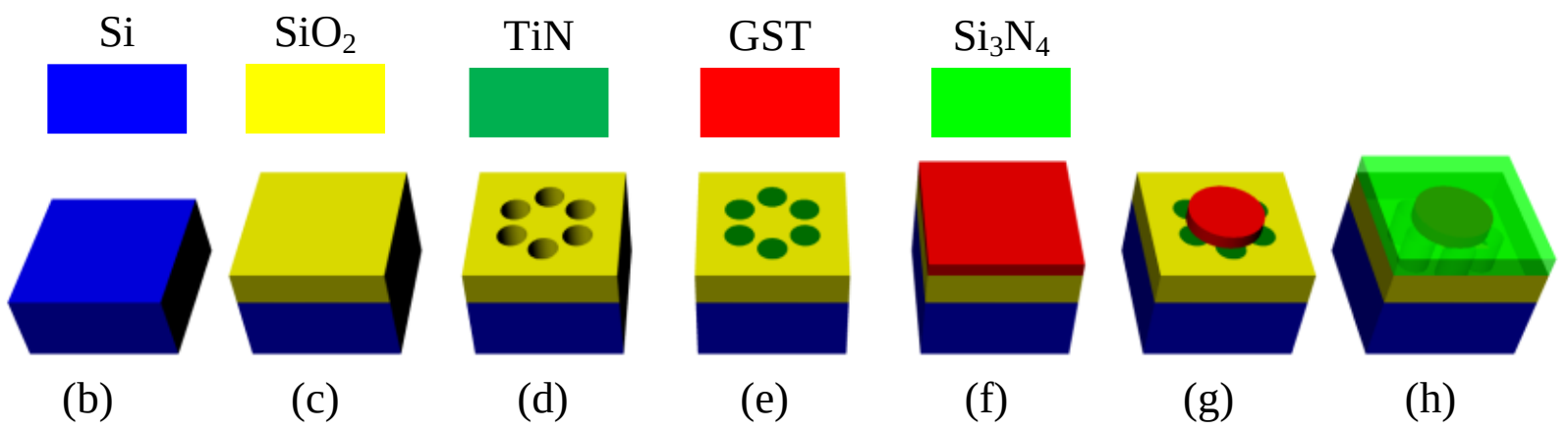

**Fig. 1.** Schematic of the (a) six-contact toggle flip-flop device, (b)-(h) fabrication process: (b)-(c) growth of $SiO_2$ on Si, (d)-(e) formation of metal contacts by etching, deposition, and planarization, (f)-(g) GST deposition and patterning, (h) deposition of $Si_3N_4$ for capping.

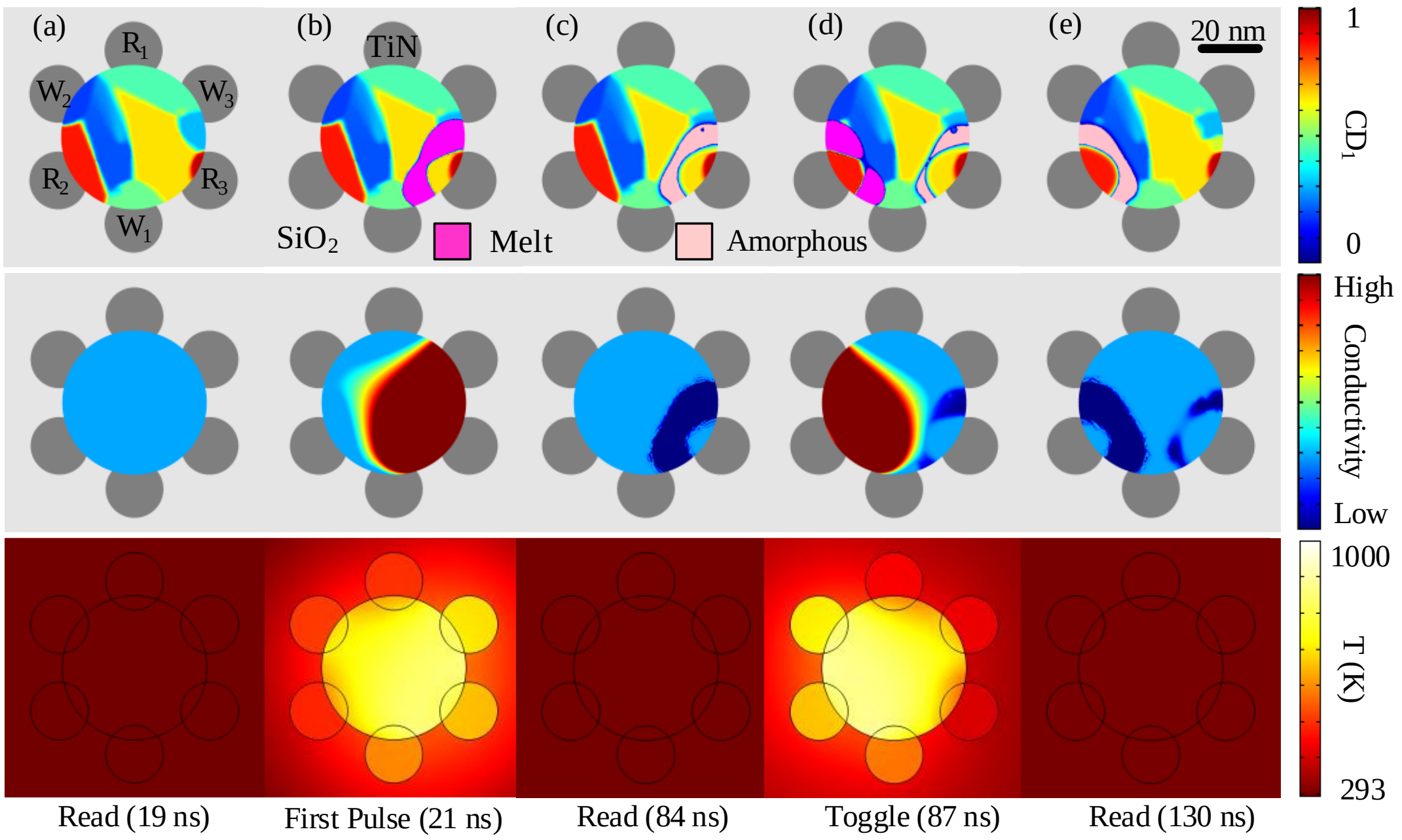

**Fig. 2.** Snapshots of electro-thermal simulation during initialization and first toggle operation. The $CD_1$ map (Top row) shows crystallinity profile of the device at different time steps. The conductivity map (center row) shows the conductivity of GST, where conductivity is lowest for amorphous (dark blue) and highest for melt (dark red). The temperature map (bottom row) shows temperature throughout the device.

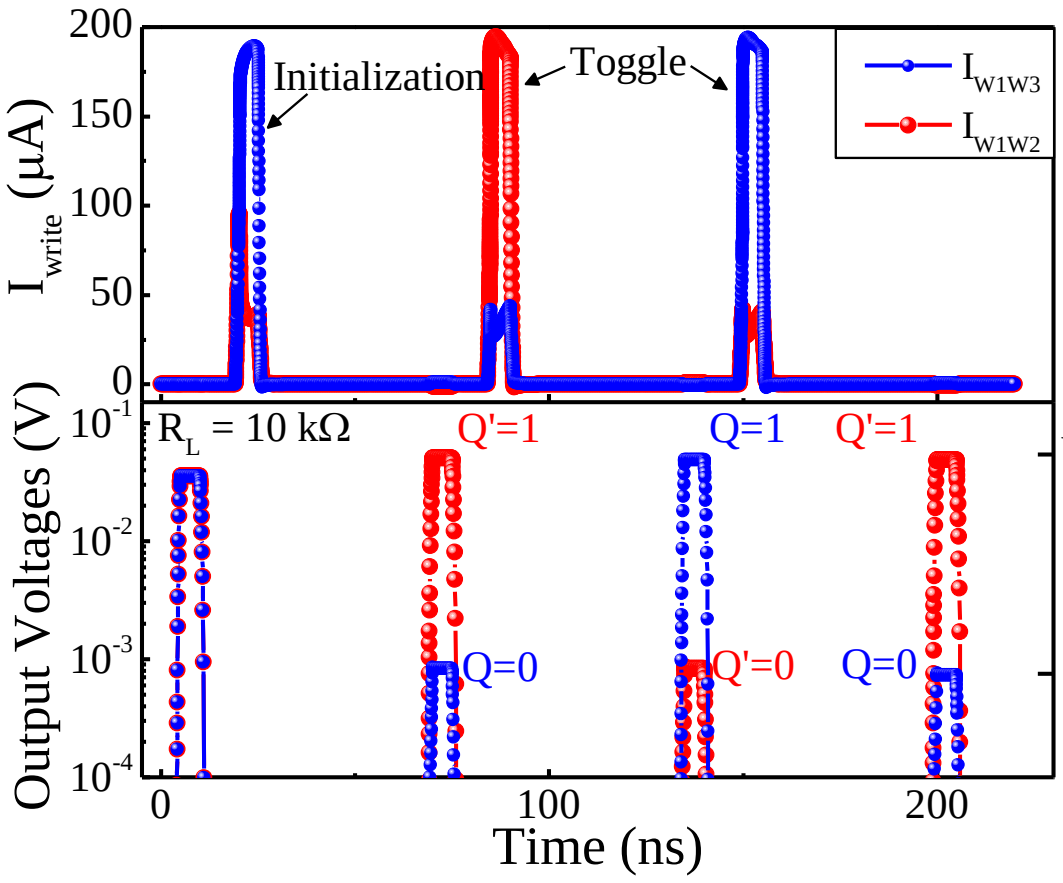

**Fig. 3.** Write currents and output voltages during initialization pulse followed by two write pulses. The output voltages (Q and Q′) toggle between $V_{low}$ and $V_{high}$ after each write pulse.

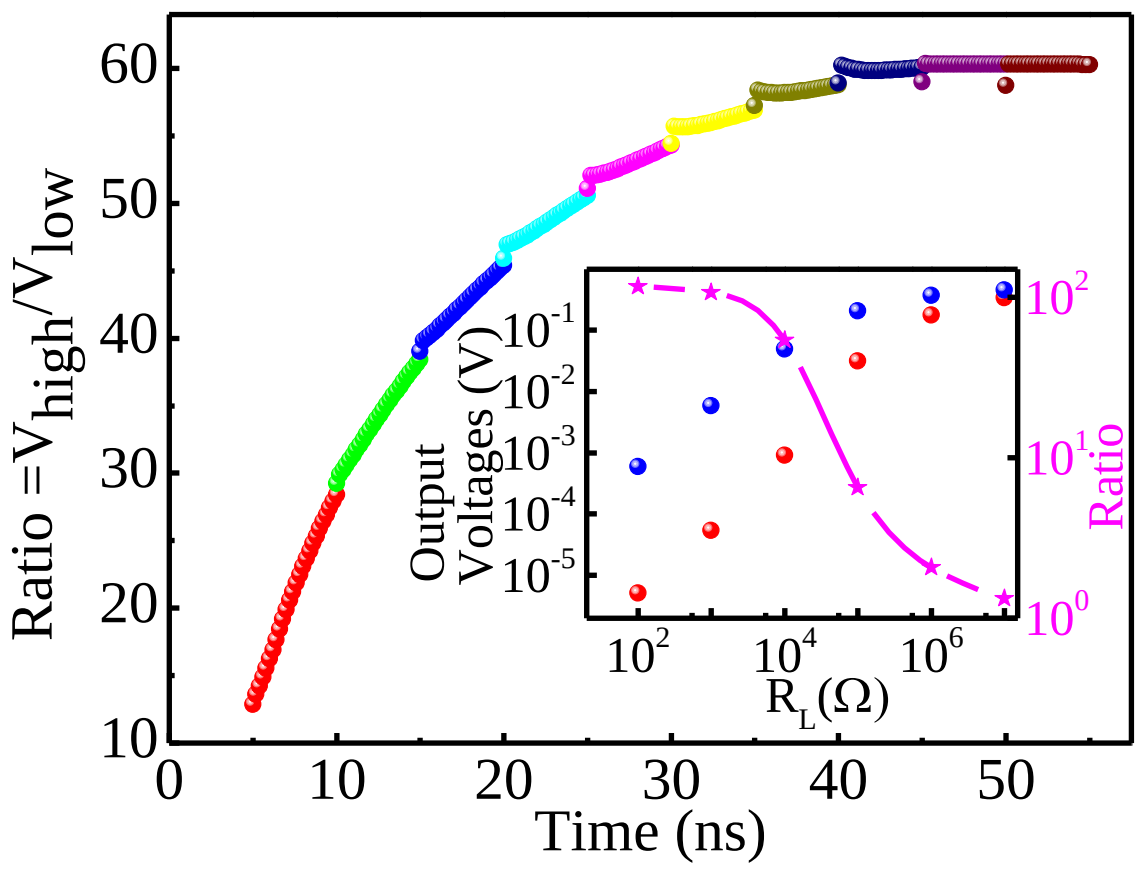

**Fig. 4.** Ratio of output voltages for read pulses (5 ns) applied at different times after termination of write pulse. Inset: $V_{high}$, $V_{low}$ (Left axis, blue and red spheres) and their ratio (right axis, pink stars) for different $R_L$ values.